\documentclass{svproc}
\usepackage{url}
\usepackage{multirow}
\usepackage{amsmath}
\usepackage{amssymb}
\usepackage{adjustbox}

\begin{document}
\mainmatter              
\title{Investigating the Role of Renewable Energies in Integrated Energy-water Nexus Planning under Uncertainty Using Fuzzy Logic}
\titlerunning{Energy-water Nexus Planning under Uncertainty}  
%
\author{Afshin Ghassemi\inst{1} \and Michael J Scott\inst{2}+
}
\authorrunning{Afshin Ghassemi and Michael J Scott} 
%
\tocauthor{Afshin Ghassemi and Michael J Scott}
\institute{University of Illinois at Chicago, Chicago IL 60607, USA,\\
\email{afshinghassemi@gmail.com}
\and
University of Illinois at Chicago, Chicago IL 60607, USA,\\
\email{mjscott@uic.edu}}

\maketitle              

\begin{abstract}
Energy and water systems are highly interconnected. Energy is required to extract, transmit, and treat water and wastewater, and water is needed for cooling energy systems. There is a rapid increase in demand for energy and water due to factors such as population and economic growth. In less than 30 years, the need for energy and water will nearly double globally. As the energy and water resources are limited, it is critical to have a sustainable energy-water nexus framework to meet these growing demands. Renewable energies provide substantial opportunities in energy-water nexuses by boosting energy and water reliability and sustainability and can be less water-intensive than conventional technologies. These resources, such as wind and solar power, do not need water inputs. As a result, they can be used as a supplement to the energy-water nexus portfolio. In this paper, renewable energies in energy-water nexus have been investigated for a range of possible scenarios. As renewable energy resources are not deterministic, fuzzy logic is used to model the uncertainty. The results show that renewable energies can significantly improve the energy-water nexus planning; however, the power grid reliability on renewable energy should be aligned with the level of systems uncertainty. The gap between the decisions extracted from the Fuzzy model and the deterministic model amplifies the importance of considering uncertainty to generate reliable decisions. 
\end{abstract}
\section{Introduction}
Energy and water are highly interconnected. Water is used to cool down power systems, and energy is used for water extraction, transmission, and treatments. Neglecting to consider the coupling between energy and water systems planning might cause disastrous consequences \cite{Gleick1993}. The meta planning system that includes energy and water planning systems at the same time is called energy-water nexus \cite{greenwade93}. Despite the growing demand for energy and water in upcoming years \cite{Stillwell2011}, generally, these resources are limited. The growing demand of common resources is predicted to be one of the main topics of conflicts in the upcoming years not only on interconnected resources but, in many other critical area (see \cite{2015JIEI...11..543E,mohamadi2020nash}). Renewable energies such as wind and solar are considered to have unlimited capacity \cite{MUHANJI2020114274}. They also do not rely on water for power generations. However, due to these power sources' uncertain nature, building a model that generates reliable decisions under uncertainty to maximize their energy-water nexus systems efficiency is essential \cite{Ghassemi2017}. To the best of our knowledge, there are a few studies that offer general mathematical energy-water nexus planning models \cite{TSOLAS2018230,ZHANG201677,en10111914,MUHANJI2020114274,WANG2018353}. Ghassemi \cite{ghassemi2019system} proposed a framework for energy-water nexus planning that minimizes the total systems cost while reducing the unwanted environmental effects.

This research aims to utilize the Fuzzy logic to provide optimal solutions that work under different uncertainty levels. Fuzzy logic has been used in numerous applications in different range of fields, to deal with uncertainty \cite{pejoo,PEYKANI2019439}. The application of the Robust optimization \cite{Ghassemi2017,ghassemi2019system}, Fuzzy logic \cite{en10111914,Chen2015,Guo2010,COELHO2016567,SUGANTHI2015585,doi:10.1111/exsy.12534} and stochastic programming \cite{babayan2005least,Weini2013,Xie2013} Have been used on a limited scale in energy-water nexus planning and in a broader capacity in the classic scientific fields. Using the Fuzzy logic makes it possible to involve the policymakers, shareholders, and managers' opinions and also scenarios-based planning into consideration. Using the Fuzzy optimization method makes it possible to find a reliable and efficient role for renewable energies in energy-water nexus systems. 

The rest of this study is organized as follows. In the next section, the energy-water systems' profile is explained. In section 3, the application of the proposed Fuzzy optimization is provided. The experiments' results are provided in section 4. Finally, the findings, results, and future work are discussed in section 5.
\section{Fuzzy Optimization Method} \label{test}
The Possibility, Necessity, and Credibility measures represent the chances of occurrence of Fuzzy events with an optimistic, pessimistic, and moderate view towards planning. Here, the conversion of Fuzzy chance constraints to their equivalent crisp for a specific confidence level is provided. In this paper, $\tilde{\mu }$ and $\gamma $ are Fuzzy and crisp numbers. It is assumed that the uncertain parameter is a Fuzzy number with trapezoidal distribution that is showed by $\tilde{\mu }\,({{\mu }_{1}},{{\mu }_{2}},{{\mu }_{3}},{{\mu }_{4}})$ while ${{\mu }_{1}}<{{\mu }_{2}}<{{\mu }_{3}}<{{\mu }_{4}}$.
\subsection{Possibility Measure}
Let the triple ($\Theta$, P($\Theta$), Poss) represent a possibility space that a universe set $\Theta$ is a non-empty set, covering all possible events and P($\Theta$) be the power set of $\Theta$. For every $A,B\in P(\Theta )$, there are non-negative numbers, $Poss\{A\}$ and $Poss\{B\}$, the possibility measure has these features:

\begin{itemize}   
\item $Poss\{\varnothing \}=0$
\item $Poss\{\Theta \}=1$
\item $if\;\; A\in P(\Theta )  \;\;\Rightarrow \;\;  0\le Poss\{A\}\le 1$
\item $Poss\{ {{\cup }_{i}}{{A}_{i}} \}=Su{{p}_{i}}(Poss\{{{A}_{i}}\})$
\item $if \;\;A,B\in P(\Theta ) \;\; and \;\; A\subseteq B \;\; \Rightarrow \;\;  Poss\{A\}\le Poss\{B\}                             \;\;  (Monotonicity)$
\item $if \;\;A,B\in P(\Theta ) \;\; \Rightarrow  \;\; Poss\{A\cup B\}+Poss\{A\cap B\}\le Poss\{A\}+Poss\{B\}      \\ \;\;(Subadditivity)$
\end{itemize}

Let $\tilde{\mu}$ be a trapezoidal Fuzzy variable on the possibility space $(\Theta ,P(\Theta ),Poss)$. The possibility of fuzzy events $\{\tilde{\mu }\le \gamma \}$ and $\{\tilde{\mu }\ge \gamma \}$ are as Eqs. \ref{1} and \ref{2}:

\begin{equation} \label{1}
Poss\{\tilde{\mu }\le \gamma \}=\left\{ \begin{aligned}
  & 1, \;\;\;\;\;\;\;\; \;\;\;\;\;\;if \;\;\;{{\mu }_{2}}\le \gamma ; \\ 
 & \frac{\gamma -{{\mu }_{1}}}{{{\mu }_{2}}-{{\mu }_{1}}},\;\;\;if \;\;\;{{\mu }_{1}}\le \gamma \le {{\mu }_{2}}; \\ 
 & 0,\;\;\;\;\;\;\;\; \;\;\;\;\;\; if  \;\;\;{{\mu }_{1}}\ge \gamma . \\ 
\end{aligned} \right.
\end{equation}

\begin{equation}\label{2}
Poss\{\tilde{\mu }\ge \gamma \}=\left\{ \begin{aligned}
  & 1, \;\;\;\;\;\;\;\; \;\;\;\;\;\; if\;\;\; {{\mu }_{3}}\ge \gamma ; \\ 
 & \frac{{{\mu }_{4}}-\gamma }{{{\mu }_{4}}-{{\mu }_{3}}},\;\;\;if\;\;\;{{\mu }_{3}}\le \gamma \le {{\mu }_{4}}; \\ 
 & 0, \;\;\;\;\;\;\;\;\;\;\;\;\;\;if\;\;\;{{\mu }_{4}}\le \gamma . \\ 
\end{aligned} \right.
\end{equation}
Based on the possibility measure, the Fuzzy chance constraints for the confidence level of ($\alpha$) are equal to Eqs. \ref{3} and \ref{4}:

\begin{equation}\label{3}
Poss\{\tilde{\mu }\le \gamma \}\ge \alpha \;\;\;\Leftrightarrow\;\;\; (1-\alpha ){{\mu }_{1}}+\alpha {{\mu }_{2}}\le \gamma     
\end{equation}
\begin{equation}\label{4}
Poss\{\tilde{\mu }\ge \gamma \}\ge \alpha \;\;\;\Leftrightarrow\;\;\; \alpha  {{\mu }_{3}}+(1-\alpha ) {{\mu }_{4}}\ge \gamma      
\end{equation}

\subsection{Necessity Measure}
The Necessity measure works as the dual model of the possibility measure. For every $A,B\in P(\Theta )$, the profile of the Necessity measure are presented as follows:

\begin{itemize}   
\item $Nece\{\varnothing \}=0$
\item $Nece\{\Theta \}=1$
\item $if \;\;A\in P(\Theta ) \;\; \Rightarrow  \;\; 0\le Nece\{A\}\le 1$
\item $Nece\{ {{\cap }_{i}}{{A}_{i}} \}=in{{f}_{i}}(Nece\{{{A}_{i}}\})$
\item $if\;\; A,B\in P(\Theta ) \;\; and \;\; A\subseteq B \;\; \Rightarrow \;\;  Nece\{A\}\le Nece\{B\}                             \;\;  (Monotonicity)$
\item $if \;\; A,B\in P(\Theta ) \;\; \Rightarrow \;\;  Nece\{A\cup B\}+Nece\{A\cap B\}\ge Nece\{A\}+Nece\{B\}      \;\;\\(Subadditivity)$
\item$if \;\;Poss\{A\}<1 \;\; \Rightarrow \;\;  Nece\{A\}=0$
\item$if \;\;Nece\{A\}>0 \;\; \Rightarrow \;\;  Poss\{A\}=1$
\end{itemize}
The Necessity (Nece) of Fuzzy events is as Eqs. \ref{5} and \ref{6} and the transformed versions are as Eqs. \ref{7} and \ref{8}:
\begin{equation}\label{5}
Nece\{\tilde{\mu }\le \gamma \}=\left\{ \begin{aligned}
  & 1, \;\;\;\;\;\;\;\;\;\;\;\;\;\; if \;\;\;{{\mu }_{4}}\le \gamma  ; \\ 
 & \frac{\gamma -{{\mu }_{3}}}{{{\mu }_{4}}-{{\mu }_{3}}},    \;\;\; if\;\;\; {{\mu }_{3}}\le \gamma \le {{\mu }_{4}} ; \\ 
 & 0, \;\;\;\;\;\;\;\;\;\;\;\;\;\; if \;\;\;{{\mu }_{3}}\ge \gamma  . \\ 
\end{aligned} \right.
\end{equation}

\begin{equation}\label{6}
Nece\{\tilde{\mu }\ge \gamma \}=\left\{ \begin{aligned}
  & 1, \;\;\;\;\;\;\;\;\;\;\;\;\;\;\;if \;\;\;{{\mu }_{1}}\ge \gamma  ; \\ 
 & \frac{{{\mu }_{2}}-\gamma }{{{\mu }_{2}}-{{\mu }_{1}}},    \;\;\; \; if \;\;\; {{\mu }_{1}}\le \gamma \le {{\mu }_{2}} ; \\ 
 & 0,\;\;\;\;\;\;\;\;\;\;\;\;\;\;\; if \;\;\;{{\mu }_{2}}\le \gamma  . \\ 
\end{aligned} \right.
\end{equation}
\begin{equation}\label{7}
Nece\{\tilde{\mu }\le \gamma \}\ge \alpha \;\;\; \Leftrightarrow \;\;\;(1-\alpha ) {{\mu }_{3}}+\alpha  {{\mu }_{4}}\le \gamma 
\end{equation}
\begin{equation}\label{8}
Nece\{\tilde{\mu }\ge \gamma \}\ge \alpha \;\;\;\Leftrightarrow\;\;\; \alpha  {{\mu }_{1}}+(1-\alpha ) {{\mu }_{2}}\ge \gamma 
\end{equation}

\subsection{Credibility Measure}
Finally, the Credibility measure (Cred) of event \{A\} is defined as the average of its Possibility (Poss) and Necessity (Nece) measures as: \[Cr\{A\}=\frac{1}{2}(Poss\{A\}+Nece\{A\}).\]
The profile of the Credibility measure are presented as follows:
\begin{itemize}   
\item $Cred\{\varnothing \}=0$
\item $Nece\{\Theta  \}=1$
\item $if \;\; A\in P(\Theta \ )\;  \Rightarrow \;  0\le Cred\{A\}\le 1$
\item $if \;\;{{A}_{i}}\in P(\Theta \ ) \;\; and \;\; Su{{p}_{i}}(Cred\{{{A}_{i}}\})<0.5  \Rightarrow   Cred\{ {{\cup }_{i}}{{A}_{i}} \}=Su{{p}_{i}}(Cred\{{{A}_{i}}\})$
\item $if \;\;A\in P(\Theta \ )  \Rightarrow   Cred\{A\}+Cred\{{{A}^{C}}\}=1                                                   \;\;\;(Self-Duality)$
\item $if\;\; A,B\in P(\Theta \ ) \;\; and \;\; A\subseteq B  \Rightarrow   Cred\{A\}\le Cred\{B\}                             \;\;\;      (Monotonicity)$
\item$if \;\;A,B\in P(\Theta \ )  \Rightarrow   Cred\{A\cup B\}\le Cred\{A\}+Cred\{B\}                                    \;\;\;(Subadditivity)$
\item$Poss\{A\}\ge Cred\{A\}\ge Nece\{A\}$
\end{itemize}
Based on the Credibility measure, transforming the Fuzzy chance constraints for the confidence level of ($\alpha$) is as Eqs. \ref{9} and \ref{10}:
\begin{equation}\label{9}
Cred\{\tilde{\mu }\le \gamma \}=\left\{ \begin{aligned}
  & 0, \;\;\;\;\;\;\;\;\;\;\;\;\;\;\;\;\;\;\; if \;\;\;{{\mu }_{1}}\ge \gamma  ; \\ 
 & \frac{\gamma -{{\mu }_{1}}}{2({{\mu }_{2}}-{{\mu }_{1}})},   \;\;\;    if \;\;\; {{\mu }_{1}}\le \gamma \le {{\mu }_{2}} ; \\ 
 & \frac{1}{2}, \;\;\;\;\;\;\;\;\;\;\;\;\;\;\; \;\;\;if \;\;\;{{\mu }_{2}}\le \gamma \le {{\mu }_{3}} ; \\ 
 & \frac{\gamma -2{{\mu }_{3}}+{{\mu }_{4}}}{2({{\mu }_{4}}-{{\mu }_{3}})}, if \;\;\; {{\mu }_{3}}\le \gamma \le {{\mu }_{4}} ; \\ 
 & 1, \;\;\;\;\;\;\;\;\;\;\;\;\;\;\;\;\;\;\; if \;\;\;                 {{\mu }_{4}}\le \gamma  . \\ 
\end{aligned} \right.
\end{equation}
\begin{equation}\label{10}
Cred\{\tilde{\mu }\ge \gamma \}=\left\{ \begin{aligned}
  & 1,\;\;\;\;\;\;\;\;\;\;\;\;\;\;\;\; \;\; if \;\;\; {{\mu }_{1}}\ge \gamma  ; \\ 
 & \frac{2{{\mu }_{2}}-{{\mu }_{1}}-\gamma }{2({{\mu }_{2}}-{{\mu }_{1}})},       if  \;\;\;{{\mu }_{1}}\le \gamma \le {{\mu }_{2}} ; \\ 
 & \frac{1}{2}, \;\;\;\;\;\;\;\;\;\;\;\;\;\;\;\;\; if  \;\;\;{{\mu }_{2}}\le \gamma \le {{\mu }_{3}} ; \\ 
 & \frac{{{\mu }_{4}}-\gamma }{2({{\mu }_{4}}-{{\mu }_{3}})},    \;\;\; if  \;\;\; {{\mu }_{3}}\le \gamma \le {{\mu }_{4}} ; \\ 
 & 0,\;\;\;\;\;\;\;\;\;\;\;\;\;\;\;\;\; if \;\;\;{{\mu }_{4}}\le \gamma  . \\ 
\end{aligned} \right.
\end{equation}
The briefed versions are presented as Eqs. \ref{11} and \ref{12}:
\begin{equation}\label{11}
Cred\{\tilde{\mu }\le \gamma \}\ge \alpha \;\;\; \Leftrightarrow \;\;\; \left\{ \begin{aligned}
  & (2-2\alpha ) {{\mu }_{3}}+(2\alpha -1) {{\mu }_{4}}\le \gamma \;\;\;if \;\;\;\alpha >0.5 ; \\ 
 & (1-2\alpha ) {{\mu }_{1}}+2\alpha  {{\mu }_{2}}\le \gamma \;\;\;\;\;\;\;\;\;\;\;\; if \;\;\;\alpha \le 0.5 . \\ 
\end{aligned} \right.
\end{equation}
\begin{equation}\label{12}
Cred\{\tilde{\mu }\ge \gamma \}\ge \alpha \;\;\; \Leftrightarrow \;\;\; \left\{ \begin{aligned}
  & (2\alpha -1) {{\mu }_{1}}+(2-2\alpha ) {{\mu }_{2}}\ge \gamma \;\;\; if \;\;\;\alpha >0.5 ; \\ 
 & 2\alpha  {{\mu }_{3}}+(1-2\alpha ) {{\mu }_{4}}\ge \gamma \;\;\;\;\;\;\;\;\;\;\;\; if \;\;\;\alpha \le 0.5 . \\ 
\end{aligned} \right.
\end{equation}
\section{Energy-water Nexus Profile}
The deterministic energy-water nexus model and the parameters such as wind generations are the same as the author's previous work \cite{ghassemi2019system}. The planning horizon is 7$\times$24, and the model generates online hourly decisions. The final model is a Fuzzy, multi-period, linear mixed-integer optimization problem. The energy-water nexus model on the water side includes water extraction, transmission, treatment, and all the associated costs. On the power side, it covers thermal power systems, such as wind power and batteries. The model also has two penalty costs; (1) Unmet demand and (2) Loss of the polluted waste water before treatment. 
\section{Scenario Analysis}
The energy-water nexus model is analyzed using three Fuzzy approaches of Possibility (optimistic planning), Necessity (pessimistic planning), and Credibility (moderate planning) to model the uncertainty for wind generation and efficiency. Trapezoidal Fuzzy numbers are used to represent the uncertain parameters. A sensitivity analysis for the confidence intervals ($\alpha$) is performed to verify the performances of Fuzzy models. The results are illustrated in Fig. \ref{tab:costcomparison}.
It should be noted that:

    \[PI_{i,\alpha} = \frac{\text{Total systems cost of method ($i,\alpha$)}}{\text{Total systems cost of the Possibility method } (\alpha=0 )}\times{10,000}\]
where $i= Possibility, Credibility, Necessity$ and $\alpha= 0, 0.25, 0.50, 0.75, 1$. 
\begin{table}[]
\caption{Total system cost comparisons of the Possibility, Necessity, and Credibility approaches}
\label{tab:costcomparison} 
\begin{adjustbox}{width=\textwidth}
\begin{tabular}{lcccccccccc}
\hline
\multirow{2}{*}{Approach} & \multicolumn{10}{c}{Total Systems Cost (\$)}                                                           \\ \cline{2-11} 
                                    & $\alpha = 0$      & PI   & $\alpha = 0.25$      & PI   & $\alpha = 0.5$      & PI   & $\alpha = 0.75$      & PI   & $\alpha = 1$      & PI   \\ \hline
Possibility                & 3151620000 & 1    & 3151972981 & 1.12 & 3152013953 & 1.25 & 3152051772 & 1.37 & 3152121108 & 1.59 \\
Credibility                 & 3152221959 & 1.91 & 3152253476 & 2.01 & 3152316508 & 2.21 & 3152373237 & 2.39 & 3152442573 & 2.61 \\
Necessity                 & 3152404753 & 2.49 & 3152445724 & 2.62 & 3152474089 & 2.71 & 3152518212 & 2.85 & 3152556031 & 2.97 \\ \hline
\end{tabular}
\end{adjustbox}
\end{table}

As expected, the Possibility approach with $\alpha = 0$ has the least total systems cost. With an increase of $\alpha$, the cost to generate reliable decisions increases as well. The Necessity approach yields the most costly choices, and the Credibility approach provides more practical decisions. The \textit {PI} index is provided to represent the gaps between solutions. These three methods give the decision-maker flexibility to find reliable choices while respecting the additional measures (e.g., budget constraint). 
\section{Discussion}
Increased demand for energy in the future could lead to raises in the energy system's water intensity, whereas renewable energies such as wind and solar could lower the water dependence.  Renewable energy sources, such as wind power, need a low amount of water. However, they randomness around them could be more than traditional power generation such as hydro-power. To have a reliable system and maximize the efficiency of renewable energies in energy-water nexus planning, the model should be resilient against a reasonable amount of uncertainty for the key parameters.  To this aim, three Fuzzy approaches with optimistic, pessimistic, and moderate conservative levels are introduced. A numerical sensitivity analysis for a range of confidence intervals is provided. The results show that utilizing Fuzzy logic, a more reliable system can be achieved, and a flexible range of solutions can be generated.  For future studies, the goal is to focus on developing customized robust-Fuzzy approaches that can generate reliable decisions with partial data.

%

%
%
\bibliographystyle{splncs03_unsrt}
\bibliography{author}
\end{document}